# Obstacles in Fully Automatic Program Repair: A survey

S. Amirhossein Mousavi, Donya Azizi Babani, Francesco Flammini

**Abstract**

The current article is an interdisciplinary attempt to decipher automatic program repair processes. The review is done by the manner typical to human science known as "diffraction". We attempt to spot a gap in the literature of self-healing and self-repair operations and further investigate the approaches that would enable us to tackle the problems we face. As a conclusion, we suggest a shift in the current approach to automatic program repair operations in order to attain our goals. The emphasis of this review is to achieve "full automation." Several obstacles are shortly mentioned in the current essay but the main shortage that is covered is the "overfitting" obstacle, and this particular problem is investigated in the stream that is related to "full automation" of the repair process.

1. **Introduction**

In today's world, security and dependability happen to become the most significant priority of system requirements whereby the system functions even in the presence of faults or cyber-attacks. An exhaustive review; *"Epic failures: 11 infamous software bugs"*, [1] illustrates most famous software system disasters caused by the system bug which has resulted in massive financial losses and even human lives losses. A bug or failure can cause drastic and unrecoverable losses in crucial and sensitive systems such as aviation systems, highways or railways traffic control systems etc. Debugging processes almost impose up to 50% of the development expenses of software production [2], [3]. However, by thoroughly investigating the existing literature at hand, one may encounter various definitions, explanations, and synonyms for bugs, fault, failure, defects, etc.

Consequently, the first step to further investigate the subject is to find a common, unifying understanding of that subject which leads us to a unique description. Finding this specific conceptualization is essential for further research in order to increase the reliability of the analysis.

Furthermore, a set of standard definition and presentation of dependability and security of the computing systems is provided in a magnificent work of Avizienis et al. [4] where they discussed the embedded systems' communication and the categorization of faults. Authors further discuss the threats to the dependability and security and establish a complete description of faults in the above mentioned topic. Figure 1 is adopted from this article in which faults are portrayed in a diagram. The discussion above provokes research in the subjects of bug fixing in various types of systems.

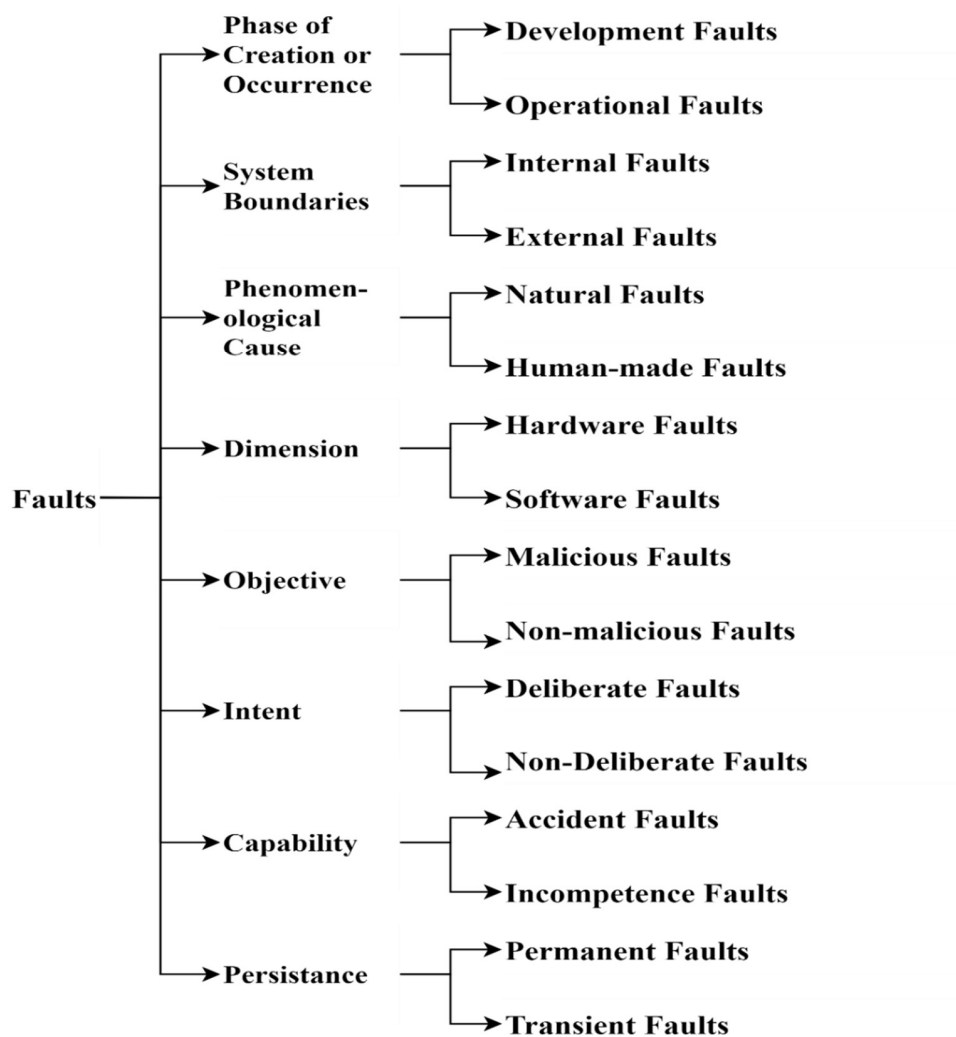

*Figure 1: Fault classification [4]*

## 1.1. Selection of literature

The current article is an interdisciplinary attempt to decipher automatic program repair processes. The review is done by the manner typical to human science known as "diffraction". Additionally, the diffractive perspective compatible with the subject of the current article is conceptualized. The novel conceptualizations are further utilized in the analysis of the theoretical aspects and the aspects relative to human relationship with the subject. Although numerous articles and research are studied in order to fully understand how the diffractive analysis methodology can become compatible to the notion of automatic program repair, we decided to merely cite the primary sources where the diffractive method is theorized [5] [6] [7].

The current article continues the study by searching keywords in Google Scholar, ACM, and IEEE with the acquired keywords in the form of a query as; `(dependability OR security OR resilien*) AND (fault OR failure OR bug OR error) AND (resilien* OR fixing OR tolerance OR recovery)`.

The given query resulted in hundreds of articles and book chapters. The next phase was to filter them according to peer-reviewed, and the number of citations by that gives validity to findings. Yet the number of findings is more than one hundred. Further, we started to look for the most thought-provoking titles that contain most of the keywords. Amongst them, we can name [8] and [9]that discuss the repair of bugs and faults in the software. Also looking at these two papers gave us the idea that automation of bug fixing needs code generating.

Further research into code generation resulted in finding a doctoral dissertation [10] in which the core of its proposed code generating techniques is based on search algorithms. In order to fully understand the impact of search algorithms on the automatic program repair, we further carefully selected the article [11] which we highly recommend as an essential read. Subsequently, along with our continuous reviewing of the selected literature, we added several additional articles. We faced a need for more in-depth research that resulted into another set of a query as; `("auto*" OR ("self-" AND ("*repair*" OR "*heal*" OR "*fix*")) AND "*overfitting*" AND "genetic*prog*"`. What will come in the following sections as a result of the review is focused on a stream which covers the shortcomings we are facing to achieve the fully automatic program repair. The emphasis is on "full" automation which means the elimination of the human interference in the whole process and secondly, the actual ability of the techniques to produce fixes for all sorts of bugs; especially the complex bugs that now are most often handled by human developers. Finally, the list of the used literature is provided in the reference list at the bottom of the current paper. However, various great papers were excluded in order to maintain the focus of the paper. Future research can cover more aspects and dig deep into the possible solutions as well.

2. **Methodology**

There exist numerous reviews on automatic software repair, wherefore we aim to pursue a new methodology that enables us to deduce new slant of the relevant issues. The analysis of the dynamic shaped between human and technology is intricate, and this intricacy is increasing by the inflating complexity of this relationship. While most contextual reviews use critical *reflection* method, the current article adopts the *diffraction* method suggested by Barad [5] in analyzing how entities interact while noticing both human and non-human materials. Such contact between two entities is called *intra-action.* An *intra-action* is when two entities come into contact and by their entanglement, a third new entity is produced [5]. This third entity may be a new understanding while reading theories through each other or a new materialization that did not exist prior to that specific intra-action. The intra-action is pivotal in the diffractive analysis because as Barad [5] argues, the production of knowledge does not happen in isolation but in between entities.

Diffraction is a metaphor with tacit referral to physics. As indicated by Barad [5] diffraction in physics refers to "the way waves combine when they overlap and the apparent bending and spreading out of waves when they encounter an obstruction". Hence, a methodology adopting such a metaphor is an inquiry of differences while accounting entanglements of the entities [5]. Barad [5] uses Niels Bohr's theories to clarify that it is not possible to separate human and non-human because the reality is the relationship between the observer and the subject. Hence, diffraction is the most appropriate metaphor for the theoretical mapping tool Barad [5] was suggesting. Diffraction stands as the opposite of reflection. As articulated first time by Haraway [6] [7] the traditional reflective scientific analysis reproduces the same as it is already given, where diffractive thinking proceeds beyond this by shedding light on the differences and locating where the effects of those differences are problematic.

Barad [5] and Haraway [7] believe that the traditional reflection displace the same concepts but it is reproducing the 'same' that does not make any difference and do not produce new meanings where diffraction is another kind of critical thinking, mapping differences and their effects, interferences and interactions with the aim of producing a new meaning. Diffractive methodology is detail oriented, irrational and dynamically theorizing. Diffraction as a theoretical mapping tool is used to broaden our horizons by acknowledging the intra-activity of matters with unclear boundaries. This means that we do not view from the perspective where humans have the agency to act upon non-human matter but it is the study of the phenomenon that comes to life via the relationship between human and non-human and the agency is bestowed to the dynamic and not the pre-existent parties alone.

The ontological difference between traditional reflective scientific review and diffraction is that when reflecting, humans and other entities are viewed with discrete boundaries and substance while in diffractive analysis as purported by Barad [5] entities whether human or non-human are viewed with fluid boundaries and the reality of substance is agential meaning that a particular phenomenon emerges by the intra-action between entities whereby new boundaries are created. Diffractive analysis enables us to understand what new substance is produced while entities at hand intra-act. The diffractive analysis usually discovers the new knowledge by two procedures. First, the procedure mostly seen in contextual analysis and theoretical work which is often referred to as *reading concepts through each other* [5] and secondly, the analysis of the intra-activities between humans and non-humans involved in the subject of the study known as the *experimenting with a diffractive apparatus*.

The current article suggests that the incorporation of the diffractive method in the subject of automatic program repair has both facets of diffractive reading which includes the level of reading concepts through each other, and the level of experimenting with a diffractive apparatus which includes the analysis with the focus on the intra-action of the human developer and the process of debugging in the road of semi to full automation of program repair. The first level includes two procedures. Namely, the procedure of 'reading concepts through each other' and the procedure of 'contextual analysis of the codes'. On the level of experimenting with a diffractive apparatus, the intra-activities involved in the process of automatic program repair are conceptualized and further discussed.

## 3. Reading concepts through each other

In this section, we discuss the two levels of automatic failure fixing. We select one of the very important mis-conceptualizations in the literature which is the "self"; conveying the meaning of full automation.

### 3.1. Differentiation of automatic fixing

There exist two levels of automatic fixing regarding program failures. First, *runtime* level in which software healing resolution detects the failure and restores the operation to the normal level of function [12], [13], [14]. Second, *source code* level in which software repairing resolution detects the failure, localizes it and fixes the fault [15], [16], [17]. As respects, if the human supervision is eliminated from the process, the automatic software healing is termed as *self-healing,* and the automatic software repairing is termed as *self-repairing*. In other respects, the healing process consists of two levels [18]. To be specified, the *healing* level in which the healing operation carried out in order to prevent or lessen the failure and the *verification* level which is not always included in the healing process but if it is, it makes sure that the application is running as required after the healing [18]. Then if the application fails to run adequately, then the healing process repeats [19], [20]. The *healing* process ensures two possibilities in which we may face with the binary of *successful execution-faulty program* or the binary of *failed execution-faulty program* which demonstrates that the program is still faulty after the healing process whether the application runs appropriately or not [18].

On the other hand, the *repairing* process consists of three levels. To be specific, the *localization* level in which the fix-required locations are spotted. Then, in the *fix* level, the required fixes in the spotted code locations generate and finally, in the *verification* level it is confirmed whether the software is repaired or not. To be precise, the main difference between the healing process and the repairing process is where the healing process focuses on preventing failures, but the repairing process generates fixes by collecting observations from several failed and successful executions in order to be able to fix the code source of the program [18].

### 3.2. "Self"-repair and literature confusion

In a different manner, Salehi and Tahvildari [21] postulate that self-healing operations are counted of the *self-adaptive* processes. They further suggest that self-healing operation is the incorporated process including *self-diagnosing* and *self-repair* hence this process encompasses the ability to identify the fault and recover the execution. Categorically, subsuming the self-repair operation whereby the code-source is adjusted; under the self-healing processes in which the code is untouched is irreconcilable [12]. Self-healing operations foremost stem from *fault-tolerant* and *self-stabilizing* systems in which Kephart and Chess [22] inscribed that self-healing operations comport deployment of *availability, survivability, maintainability,* and *reliability* of the system. In a similar vein, Ghosh [23] suggests that failure classification in self-healing systems is similar to fault classification in *fault-tolerant systems*. Classification of the fault is substantial hence the recovery strategies activates only in existence of the classified fault. Whether the fault is affecting a single unit or more, because of the interdependency of the units, it may cause more and more failures in the system [12]. Other scholars, on the other hand, proposed different classifications of faults in self-healing systems. For instance, Kopetz [24] suggests that ramification of the faults is related to their value or timing. One important fault class is the two-faced failure type in which the system can merely identify it by $3k + 1$ components ($k$ is the amount of tolerated failures) [28]. Another classification can be considered as the amount of failure in a specific time interval. Although the different classification methodologies can be vague and confusing, they can be implemented based on the needs and requirements of a particular program [28].

Psaier and Dustdar [28] conclude that self-healing systems contain *fault-tolerant* operations along with *self-stabilizing* abilities and finally, *survivable system* operations. The ideal is that the processes can operate automatically, but in case of severe failure, still, human supervision is needed in the proposed self-healing systems which are incompatible with the notion of "self" healing. This contradiction of human invasion in the allegedly automatic system is justified by the requirements that are unknown in the stage of production and can only appear over time [28], [34]. However, the current article suggests that in most cases we can prevent the imminent failures by *test scenarios* or *test-driven implementation* or *exception handling* in the production stage. Second justification mentioned by Psaier and Dustdar [28] is the fact that the self-healing process will continue repeating itself until no further action can be applied.

The second justification is basically a part of the definition of the self-healing function mentioned by Gazzola et al. [18] and is the phase the system needs to operate the *self-repair* stage where the code-source of the faulty program gets corrected. The two given justifications lead to the suggestion by Kephart and Chess [35] in parallel with Ghosh et al. [13] and Salehi and Tahvildari [34] in which they impart the necessary human assistance in the self-healing process which results in *assisted-healing* instead of an automation circumstance. This article suggests that the self-healing system as an actual automatic operation is attainable if the proper preparation in the production stage takes place, such as the above mentioned techniques of test-scenarios, test-driven implementation, and exception handling. However, one may argue that this may increase the time and required resources of the production stage, but indeed, it will still range from the human invasion of semi-automatic proposed healing operations.

However, Psaier and Dustdar [28] articulate that self-healing research is rooted in *recovery-oriented* operations which encompass the *stabilizing* system as a part of it along with other strategies for repairing and preventing further faults. Figure 2 is adopted from Psaier and Dustdari's work to illustrate the self-healing loop.

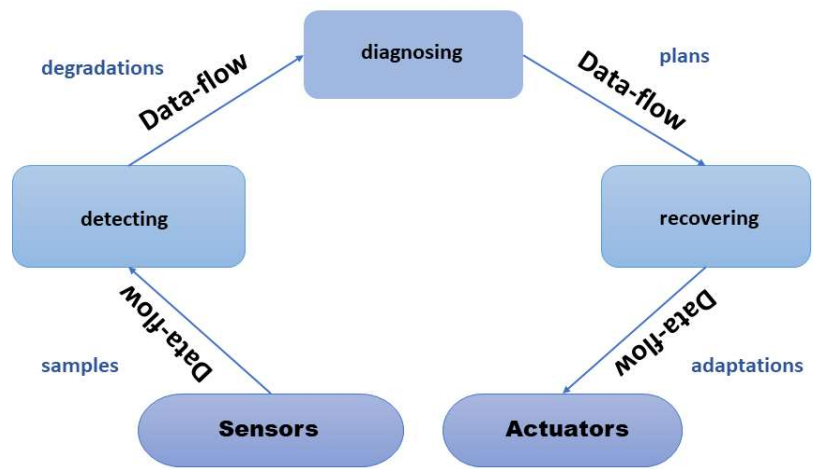

*Figure 2: self-healing loop [28]*

The degraded state propounded by Ghosh et al. [29] portrays the state in which the self-healing operations lead the system to function on a still acceptable level which depicts that systems do not suddenly stop operating but will continue the operations with a significantly lower level of performance. The given explanation indicated that the self-healing process does not correct the faulty program but buys time in order to prevent the loss of valuable data. Therefore, contrarily to what Salehi and Tahvildari [34] suggested, the self-repair system is not included in the self-healing processes as a part of the operation or as a subsume.

### 3.3. "self"-repair diagnosis

Under other conditions, *self-repair* consists of distinct phases such as *localization*. Dissimilar to self-healing, self-repair composite algorithms whereby the location that the fix must be applied can be identified. The *fault localization* has two facets. In one, the faulty statement can be spotted in order to enable the self-repair system to correct the faulty statement, and in another facet, the localization technique, namely the *fix locus localization* finds the statements that are influenced by the existing fault [18], [25]. Whether the spotted influenced statements are inherently faulty themselves or not, in some cases by adjusting or changing them, the faulty program gets corrected [18]. The most common technique used for this issue is the *Spectrum-Based Fault Localization (SBFL)* [26]. As a result, a list is provided in rank order for statements in which it is illustrated how many *behavioral tests* a statement has passed and how many tests has it failed [27], [28], [29]. When a statement fails more tests in comparison with the tests it has passed, it demonstrates that this statement has a higher potential to be faulty.

Regarding the *algorithms* (HELP! Maybe the machine learning algorithms can be implemented in this stage) for the automatic generation in fixes, it is noteworthy to say that all algorithms *approximate* the automatic generating issue [18]. According to Gazzalo et. Al [18], the problem of fix generating, can be approached in a way that by resolving the $P_{repair}$ we can solve the fix generating issue. However, a *plausible solution* is needed to be generated in order to resolve the $P_{repair}$. This plausible solution, namely the $S_{repair}$ is mainly depended on the way $P_{repair}$ is defined. $P_{repair}$ can be addressed by the approach *generate-and-validate* in which a space $S_{repair}$ is generated to resolve the $P_{repair}$ by various produced elements containing elements that can actually resolve the $P_{repair}$ and elements which can not. Another vein to address the $P_{repair}$ is *semantics-driven/correct-by construction* approaches in which $S_{repair}$ is not produced. Instead, the solution is encoded as a part of the $P_{repair}$, whether in a restricted way or in an abstract implicit way. The solution which is found by this approach has the highest possibility to solve our $P_{repair}$ problem. However, the solution is still reported to the human developer who is responsible for checking the accuracy of the fix. The extent that self-repair systems operate automatically is admirable, but one may argue that the repair process cannot be completed without human assistance. Hence, it is not yet a fully "self" repair operation [18]. The four following tables [18] are adopted to portray the summary of the techniques and operators in self-repairing systems.

| Strategy | Fault Model | Change Model | Techniques |
|---|---|---|---|
| Search Base | General | AST reuse/insert/delete | GenProg, Marriagent, RSRepair |
| | | AST modifications | JAFF |
| | | Operator Replacement, Variable Name Replacement | pyEDB, MUL-APR, CASC |
| Brute-force | General | Operator Replacement, Condition Negation | Debroy and Wong |
| | | Method call insertion/deletion | PACHIKA |
| | | Functionality deletion | KALI |
| | | AST reuse/insert/delete | AE |

*Table 1: Generate and validate techniques - Automatic operators [18]*

| Template Type | Strategy | Fault Model | Change Model | Technique |
|---|---|---|---|---|
| Pre-defined | Search-Based | Concurrency Fault | Synchronized region manipulation | ARC |
| | Brute-force | General | Code transformation templates | AutoFix-E, AutoFix-E2 |
| | | | Condition change, Variable Value Change | SPR, Prophet |
| | | Buffer Overflow | Buffer Manipulation, Function replacement | PASAN, AutoPAG |
| Example Based | Search-Based | General | Code transformation templates, reuse of statements in the same application | History-Driven Repair |
| | | | Code transformation templates | RAR, Relifix |
| | Brute-force | General | Code transformation templates | R2Fix |
| | | Buffer Overflow | Insertion of code fragments from donor program in the code under repair | CodePhage |

*Table 2: Generate and validate techniques, template-based change operators[18]*

| Fault Model | Change Model | Techniques |
|---|---|---|
| General | Synthesis of new expressions | SemFix, DirectFix, Angelix, SearchRepair |
| Wrong conditions and missing pre-conditions | Condition change and if condition insertion | NOPOL, Infinitel, DynaMoth |
| Concurrency faults | Critical region manipulation, parallelization keyword move | AFix, CFix, HFix, Surendan et al. Axis, Grail, Lin et al., DFixer. |
| HTML generation fault | String modification | PHPQuickFix and PHPRepair |
| String sanitization | Insertion of checks, string modifications | SemRep, Yu et al. |
| Access control violations | Insertion of role checks | FixMeUp |
| Memory Leaks | Insertion of free() statements | LeakFix |

*Table 3: Semantic-drive techniques[18]*

| Fault Model | Techniques |
|---|---|
| General | BugFix, MintHint, Logozzo et al. QACrashFix |
| Security Fault | BovInspector, Abadi et al. CDRep |
| Data type misuses | Coker et al. , Malik et al. |
| Concurrency Fault | ConcBugAssist |
| Performance Fault | Selakovic et al. , CARAMEL |

*Table 4: Fix-recommender techniques[18]*

### 3.4. Faults and automatic program repair

However, the question of how to increase the effectiveness of a fix should be considered in another paper. Although, as mentioned by Gazzalo [18], in an investigation by Fast et al. [30] they attempt to minimize the amount of tests that are required for evaluating a fix's efficiency by using randomly assigned sets of tests that are part of a test suite. Then if the fix is able to pass the subset, then the whole test suite can be run [30]. Fast et al. [30] disclose that this method can make GenProg 81% faster. It has been revealed that the most crucial part of the generate-and-validate techniques is the sampling of the faulty behaviors. Also, the fix that targets the fault needs to be in the search space of the given technique for the system to be able to provide the most efficient repair [15], [17], [31], [32].

Alternatively, self-healing operations have been employed in various areas in regard to extending the functionality lifetime in the existence of a fault without generating a fix to correct the faulty program. For instance, *embedded systems* used in critical applications largely employ fault-tolerance in order to maintain their control over the environment without total shut down. The proposed solution for embedded systems depicts that the whole system should be included in the recovery process instead of a single element. There should be several connections made between the resources and tasks in the design stage which makes it possible to have multiple tasks running at the same time, some as idle which can compensate for other components in the presence of a fault [12].

Comparatively, as another example, *operating systems* face two sorts of faults. *Soft faults* which are recoverable incidents which the system can continue operating and *hard faults* which result in system restart and loss of data. The self-healing systems are to avoid the hard faults which can be described as *micro-reboots* on the faulty units without disrupting the whole reboot tree [33].Another approach in operating systems is the *predictive* self-healing suggested by Shapiro [34] in which a fault is foreseen by a clever *diagnosis engine* in order to enable substantial *degradation*. This proactive action is similar to what happens in the Spectrum-Based Fault Localization (SBFL) in self-repair systems in a way that it foreshadows the fault and prevents it, but the difference is that in self-repair systems the incident is applied in the production stage but in the self-healing system the process happens in the run-time.

## 3.5. Obstacles and an example: GenProg VS Relifix

In a similar vein, research suggests that the automatic bug fixing processes are cheap. In a study cited by Gazzalo et al. [18] they have found that the cost of automatic repair for AE is 4.40$ and the same circumstance for GenProg costs 14.78$ which indicates the low expenses of automatic processes. Contrarily, what should have been measured is the cumulative amount of costs and expenses for the combination of automatic fix generating and the human (developer) supervision who is there to evaluate and approve the fixes that are to be implemented.

Two other crucial obstacles in automatic bug fixing research are first, how would a system choose the accurate model of the change that is required in a specific situation without human supervision and secondly, how would a system maintain itself when it is dealing with automatic changes instead of controlled human supervised changes [18]. The abovementioned obstacles demonstrate that numerous strategy changes are required to achieve a totally automatic bug fixing system. The subjects are interdependent, and no process eventuates in isolation. A holistic view is required to help systems evolve toward full automation.

As an instance, GenProg was developed in the manner of continuing the bug fixing process as an automatic process [35], [36]. GenProg uses *genetic programming* and is able to generate fixes for a various number of faults in legacy software [37]. Although, GenProg faces some obstacles to achieve its full purpose. For example, lack of competency benchmark bugs and programs that are able to portray real-world systems and problems that a human developer rate as crucial and the defects should be generalizable [37]. According to Le Goues et al., [37] current benchmark suites such as SPEC or Siemens [38] lack such competencies. Other challenges for GenProg are namely, *scalability*; meaning the time and size of the code that can be handled and *generality*; meaning the number of sorts of faults that can be addressed. As mentioned before, the most important challenge is how credible the results would be evaluated by the developer hence the automatic loop still ends with the human interaction [36], [37].

On the other hand, Hwei Tan and Roychoudhury [39] compare GenProg in a study with Relifix. They articulate that Relifix is able to fix bugs approximately five times more than GenProg in their experiment while having a lower chance of introducing new regression comparing to GenProg. Figure 3 is adopted to portray Relifix's workflow.

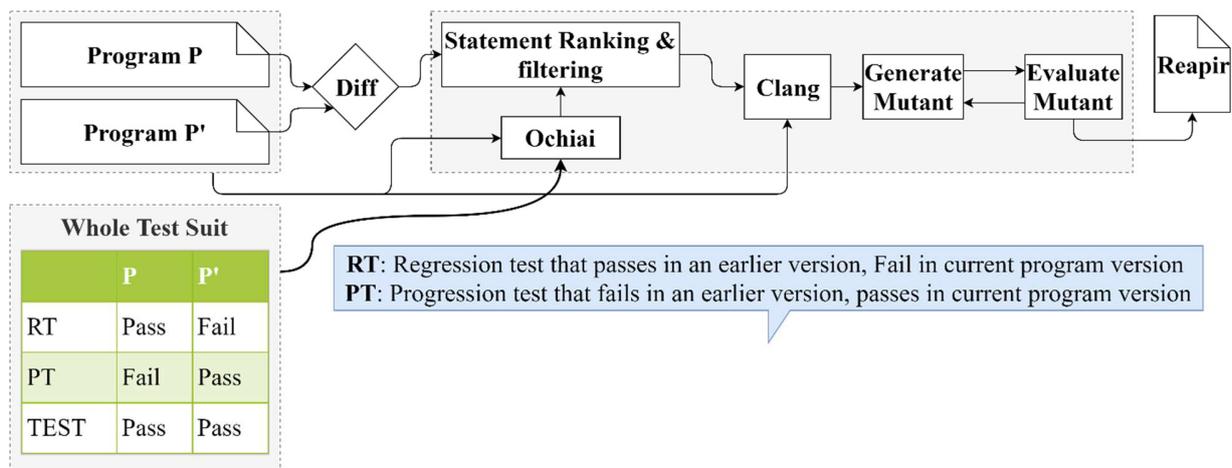

*Figure 3: the Relifix workflow [48]*

In the abovementioned study, Hwei Tan and Roychoudhury [39] suggest that relifix as a self-repair approach can contribute as a solution for self-adaptive systems by repairing regression. They conduct a comparison between GenProg and relifix in which they used a set of contextual operators including 73 real-world software regressions. They suggest that further comparisons are required in order to draw a better conclusion, but the evidence of this study indicated that relifix is a stronger technique compared to GenProg [39]. However, there is no mention of any attempt in extending the automation process to eliminate the human supervision for validating the fix before implementing in this study. This aspect is typical and repetitive among other examples as well. There is a need for developing a functionality that can evaluate the accuracy of the generated fix before implementation similar to human supervision, such as algorithms that can learn from examples of human-accepted generated fixes.

### 3.6. Time effective example

Le et al., [40] argue that existing generate-and-validate and test case driven automatic program repair techniques are not time effective. They impart that the given techniques are incompatible with the real-world bugs and most often fail in the bug fixing process, spending excessive time and resources. They track that most existing techniques fail even given 12 hours of computation in a multi-core cloud sp Hence, Le et al., [40] propose a new technique they called *history driven*. History driven technique's purpose is to mine the knowledge from bug fixing history during the development process of various sets of repositories of software. This process is done with the aim of automatically exploiting bug fixing patterns in the real-world in order to help the identification of the most relevant fix candidates. Le et al., [40] articulate that the history driven program repair work is done mainly in three phases; namely, *bug fix history extraction, bug fix history mining, and bug fix generation.* Le et al., [40] utilized existing mutation operators for generating fix candidates and then prioritized the fix candidates that matched the historical pattern in the random search process. The spotlight here is a list of ranked fix candidates that should further pass all test cases. Consequently, they succeeded to produce good-quality fixes compared to the baseline with the average time of 20 minutes. Although this technique is time-effective, it does not solve any of the issues related to semi automation of the program repair process.

## 4. Discussion

The current article would like to suggest that the most visible gap in the given literature is the complete automation operation. The automatic loop is always broken, in both healing operation and repair operation by human supervision and in this section, we attempt to analyze possibilities that will increase the chance of achieving full automation.

### 4.1. Test-suites

Yi et al. [41] impart bug fixing in a fully automatic manner is an ideal circumstance in which we can decrease the cost of programming. Therefore, the research focus is slowly shifting toward the factors that can reduce the amount of work needed by a human supervisor. They postulate that by increasing the quality of the generated patches, we can reduce the necessity of human supervision. Yi et al. [41] propose that by monitoring the quality of test suites, we can increase the quality of the generated patches. They conduct an investigation into the issue of using traditional test suites metrics for software testing such as statement/branch coverage or mutation score for improving the quality of the generated fixes [42], [35], [41]. They thoroughly investigated the correlation between traditional test-suite metrics and the reliability of the generated patches and concluded that the correlation exists [42], [35], [41]. To be precise, the correlation is strongest in statement coverage [41]. Therefore, the above mentioned technique can help us achieve a better quality for generated patches which can reduce the chance of human interference in automatic operation.

## 4.2. Overfitting

Regarding the overfitting problem of the patches, Mechtaev et al. [43] schematize a methodology regarding patch generating. They contend that the reason we face an overfitting problem in automatic repair programs is that we mostly bank on tests and further, the approaches concerning this issue are not equipped to generalize their fixes [44], [45], [46], [43]. In order to overcome overfitting and to be able to generalize the generated patches, Mechtaev et al. [43] propose exploitation of *reference implementation*. Figure 4 is adopted to illustrate the workflow of their proposed method.

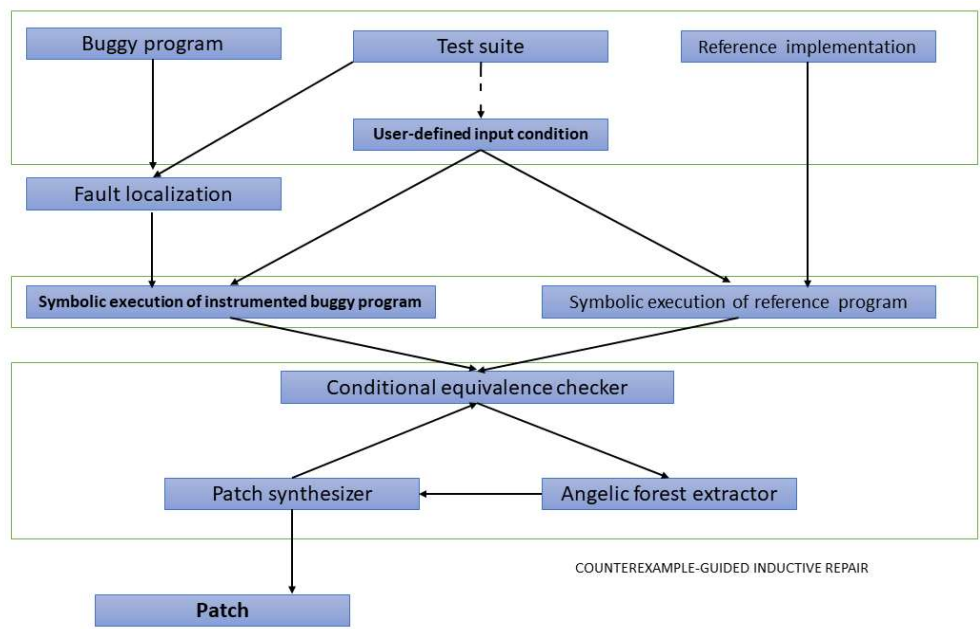

*Figure 4: workflow, reference implementation method [53]*

As apparent above, the *symbolic analysis* is to provide a specification for pre planned behavior. It is then applied in order to synthesize the fix to execute the *conditional equivalence* as portrayed in Figure 7 [43]. According to Mechtaev et al. [43] using reference implementation instead of test suites, provides us an *implicit* correctness criterion. They claim that not only their approach is more sophisticated than existing regression repair operators such as Relifix, but it is adaptable to real-world software and is able to overcome the overfitting problem. Yu et al. [47] indicate that test suites are basically input-output specifications and that is the reason they cannot make the specification for the intended behaviors which will consequently result in overfit generated patches that are incapable of generalizing beyond the given test suites. They further analyze the overfitting problem and introduce two main overfitting issues, namely, the *incomplete* *fixing* and *regression introduction* [47], [48]. The two mentioned issues respectively indicate some parts of the bug are fixed, not all, and some parts of the correct input are broken due to the generated patch. They introduce three different overfitting patches exploited from the two given issues [47]. Figure 5 is adopted to illustrate the three sorts of overfit patches.

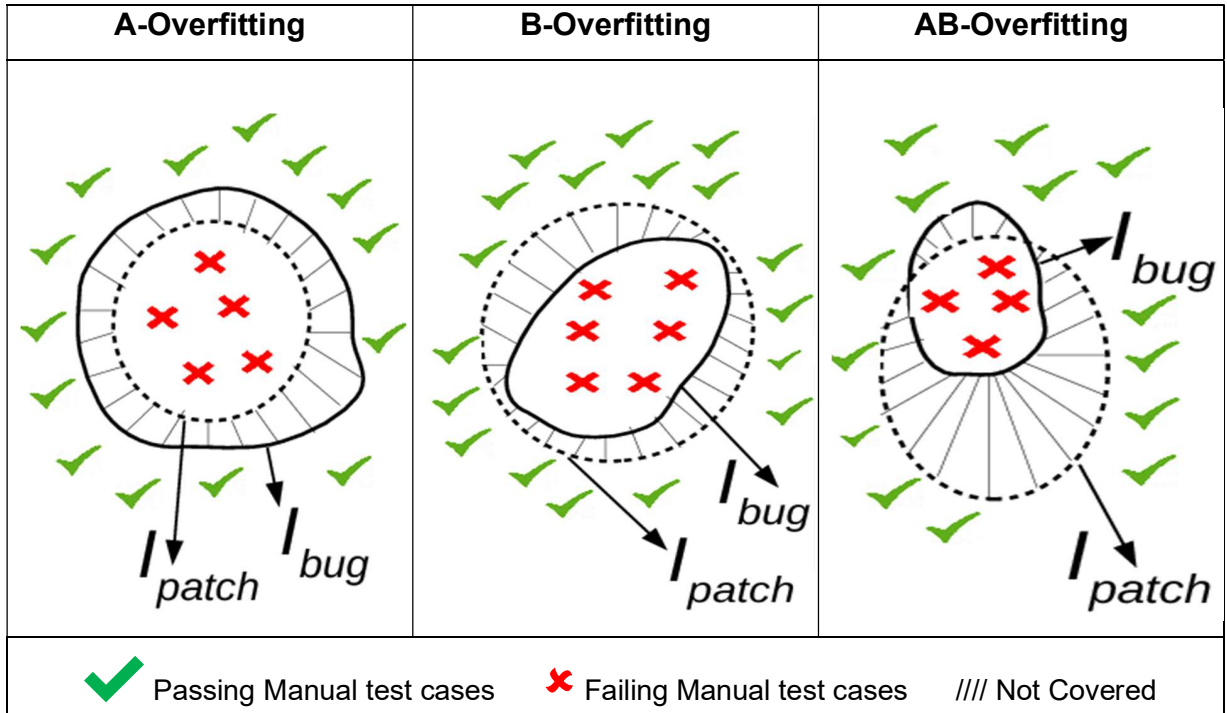

Figure 5: Three sorts of overfitting patches [57]

### 4.3. Overcoming the overfitting

Yu et al. [47] propose an approach with the aim of mitigating the overfitting problem. *UnsatGuided* is an approach for synthesis-based repair operations that automatically generates additional tests to assist repair constraint. They conclude that UnsatGuided is able to mitigate the second kind of overfitting issue; namely, the regression introduction but fails to compensate for the first issue [47], [48]. A step further, in a similar vein with overcoming the overfitting issue, Hua et al. [49] propose an on-demand technique called SketchFix. This approach tries to manifest transformation designs at *granularity* and offers small deviation to the original program [50]. The central premise of SketchFix is to decrease the *search space* by utilizing the runtime information [51], [49], [50]. The transformation is considered from a program repair to program synthesizing in which the faulty program is atomized to *sketches* at the *AST node-level granularity* [49], [51], [52]. Hua et al. [49] conclude that the new design is able to generate patches with somewhat higher quality, similar to human-made fixes.

Le et al. [53] posit that the automatic program repair techniques can be categorized into two branches; semantics-based and heuristic-based. Semantic-based utilize symbolic executions such as reference implementation [43] which further use program synthesis to enact the fix [54]. Le et al. [55] vindicate in order to overcome the overfitting problem in such approaches; we should consider using multiple synthesizing engines. Heuristic-based automatic repair systems such as Sketchfix [49] generate a vast number of candidates and use the search space to find the best fit [53]. However, both categories are still heavily relied on test suites hence the patch is evaluated as correct if it passes all the tests which lead us to the problem of overfitting [53], [56].

Xiong et al. [57] introduces a heuristic approach to establish if a generated patch is correct. They suggest that by examining the *behavior similarity* of test case executions, we are able to prevent more than half of the incorrect patches from being generated. Xiong et al. [57] indicate that passing tests behave similarly on both the original and the patched programs. In addition, they have similar behavior in the runtime and cause similar results. Therefore, we can generate a new test suite based on the similar behaviors which will be able to determine the correctness of the generated patches better [57].

Differently, concerning the discussed technique GenProg, it is noteworthy that this approach is anticipated by *Genetic Programming* (GP) [58] and is insofar practiced in different languages such as Java [17], [59], compiled code [60] and as analyzed by Oliveira [58] for C [61], [62], [15]. Olivera et al. [63] previously proposed a new s*ubpatch representation* for GenProg regarding the *crossover operators,* and they tried to achieve furtherance in their suggestion by *meta-heuristic search strategies* such as genetic programming [58]. Genetic programming automatically expands patches in order to correct the input program [58]. The automatic expansions by GP can include *functional* improvements such as debugging [61], *feature grafting* [64], [65] or the given improvements can be *quality-oriented* such as energy usage [66], [67].

Oliveira et al. [58] exploit their approach from novel research in *Search-Based Software Engineering (SBSE)* [68]. The genetic programming as described above helps the automatic operation to enforce the fix with transferring functionalities from an existing program to another [69], [65]which in return reduces the costs of debugging with reliance on test suites [70], [71], [72], [73]. Oliveira et al. [58] emphasize that the fundamental characteristic of their suggestion is the *representation* of the candidate fix as an *edit program*, on the main program. Other scholars postulate that the adaptation of the *tree-based* representation can help the programs to evolve toward the required set of qualities [15], [74].

Oliveira et al. [58] designate GenProg for their work and further explain that a patch comprises high-granularity edits which upheld three units; *edit operation, fault location* and *fix statement* whereby we can manipulate, reconstruct and perform mutation which will decrease the effectiveness of the generated repair in the search space [58]. As an alternative, they propose a new formation of the program repair representation, crossover and mutation operators. They impart that the new form should enable the transfer of the three subspaces; namely, *the operators, fault* and the *fix space* [58]. They demonstrate a lower-granularity representation that results in less limited search strategies. However, their novel representation remains linear while the subspace values are manipulated. They conclude that using genetic programming for *recombining* the above mentioned units makes the automatic repair systems more effective and evolutionary operators have a drastic positive impact on search results [58].

That having been said, Chen et al. [75] argue that the most proper way to tackle the overfitting problems is techniques that aim at the code by utilizing contracts [76]. Contacts can be exemplified by preconditions and post-conditions which will help us exploit more information in order to improve the specification of "correctness." [77]. However, Chen et al. [75] infer that common programming languages do not contain these annotations. Therefore, they present a technique called JAID [75] which is similar to described contract-based approaches [76]. JAID enables the system to generate high-quality patches, comparable to human-made, and it also decreases the overfitting problem. JAID targets Java automatic program repair systems and is highly promising [75], [78].

### 4.4. Deep learning

Scholars have been successfully utilizing deep learning algorithms in research on a variety of subjects relative to natural human language. Programming languages are not fundamentally different from human languages. There is an underlying substantial structural information distinct to programming languages [79]. The similarities between programming languages and human natural languages have been noticed by a new wave of studies in recent decades. Hindle et al., [80] analogically investigated the regularities in programming language relative to human language in order to detect the *naturalness* of the program languages. Ray et al., [81] also investigated the naturalness of the codes by considering a large corpus of bug fix commits (ca.~8,296), from 10 different Java projects. They exploited the naturalness of the buggy codes by focusing on their language statistics and their comparable fixes. They concluded that a bug has less naturalness and as the bug is being fixed, the naturalness of its code increases. These findings are pivotal for diffractive analysis since they give us a criterion for spotting the differences as we do in the natural language. The distinct similarity connecting the two fundamental structures of the languages, digital and natural, can work as the starting point.

Building up on the previous research, Gupta et al., [79] introduced an end-to-end solution to fixing programming errors automatically called *Deepfix*. Deepfix is a deep learning adaptation and although it was only examined on C language by Gupta et al., [79], they tend to also generalize the result to other programming languages. They examined Deepfix on 6971 erroneous C programs written by students where it was capable of completely fixing 27% of the errors and partially fixing 19% of the errors. Although Deepfix is evaluated on the common programming errors and not highly complex bugs, it is noted that it works without any reliance on interference for locating the fault and/or for fixing it [79]. The future of automatic program repair based on deep learning is promising.

On the other hand, compilation errors are of the most cost damaging errors in the development of software. Mesbah et al., [82] propose a novel technique called *DeepDelta* in which they utilized a deep neural network. They demonstrated that when developers change the source code error to overcome the build errors there exists patterns. Pairing failure and their fix they extracted the AST diffs which allows the DeepDelta technique to learn the above mentioned code change patterns. Consequently, DeepDelta will be able to suggest AST changes corroborated by the use of deep neural networks. Results show that DeepDelta is able to generate fixes with up to 50% accuracy while the correct fix is in the top three 85% to 87% of the time [82]. Mesbah et al., [82] indicated that they aim to expand their work in future.

Deep learning is further permeating the research on automatic program repair. Most strategies are centered on the belief that better quality of the generated fix is the solution to achieve full automation. Various scholars tend to approach this notion from different angles. For instance, White et al., [83] propound a search strategy called *DeepRepair*. They postulate that since large programs encompass the seeds or their own fix, hence it is possible to use code similarities for prioritizing and transforming statements in the codebase with the aim of generating patches. Relying on deep learning, White et al., [83] suggest that with a certain level of granularity, code fragments can be sorted by the rate of their resemblance to suspicious elements. The statements further can be transformed by the mapping of out-of-scope identifiers to corresponding identifiers. However, White et al., [83] strategies did not exceed the level of work done by redundancy-based techniques but they deem the contextual patches generated by these two different techniques are distinct. Current article suggests a mapping on these differences for future research.

By the same token, Hajipour et al., [84] pursuing a sampling based strategy, initiated a promising framework called *SampleFix*. They articulated that the key challenge to achieve full automation by the aid of deep learning, machine learning and NLP is *ambiguity*. They posit that the ambiguity is that multiple fixes may implement the same functionality and dataset is impotent in detecting the variance caused by such ambiguity [84]. Hajipour et al., [84] explicate an automatic error correction deep generative model that operates with learning the distribution of potential fixes. In order to reinforce the model to generate diverse fixes they created a Diversity-sensitive regularizer. Hajipour et al., [84] are optimistic that the results of their experiment with SampleFix indicating 61% fixes of common programming errors is a sign that this approach can be the potential basis for future advancements. Current article would like to suggest that the Diversity-sensitive regularizer should be further studied since it may have the potential to initiate a set of research toward overcoming the overfitting problem of the automatic generated fixes.

Another strategy adopted by Long and Rinard [85] resulted in a patch generation system called *Prophet*. Prophet learns a probabilistic model from previous successful human generated patches and applies the model to further prioritize automatic patches. Long and Rinard [85] posit that by learning the model parameters which are composed of identified essential characteristics of previous successful fixes, Prophet is able to apply the model after generating a search space to prioritize the most likely correct patches. This notation indicates that characteristics existing in previous successful patches follow a pattern that can be learned and further applied. This is compatible with most research initiating a learning strategy in order to advance the automatic program repair. While the similarities helped current techniques to exploit patterns, the current article suggests that by mapping the differences after exploiting the pattern we may be able to better understand what has to be done in order to achieve better results in fixing complex bugs automatically.

## 4.5. Is the current path really helping us?

To the greatest extent, the current article would like to suggest taking a step back and investigate if the automatic program repair is radically helping the developers to affect less amount of work. This is a crucial proceeding hence we should evaluate if the current manner is actually serving us in the way that we expect. In a similar vein, Motwani et al. [86] divulge that the current research regarding the assessment of automatic program repair is mostly focused on the processes that each technique goes through and the ways to increase their accuracy while there is almost no meta-analysis regarding the nature of the automatic program repair.

Motwani et al. [86] try to address a substantial question; *"are automatic repair operations fixing the bugs that are "hard" for human developers?"*. In order to gain insight regarding this issue, Motwani et al. [86] carefully defined the criteria of "hard" and then annotated two benchmarks, namely, ManyBugs and Defects4J. They carefully defined the patches' *complexity* and the *quality* of the test suites and then investigated the correlation between those characteristics and the ability of the repair technique in *patch generation* for techniques such as AE, GenProg, Kali, Nopol, Prophet, SPR, and TrpAutoRepair [86].

Discouragingly, Motwani et al. [86] found that in "hard" circumstances such as when the developer is demanded to write massive amount of code lines and/or is to edit several files and/or have numerous test suites to deal with, the automatic program repair techniques are incapable of generating the required patches [86]. However, amongst all, the Java techniques such as above mentioned JAID [75] are more capable of generating *high-priority* patches [86], [45]. Additionally, Motwani et al. [86] revealed that there is no correlation between the *time* that a human developer takes to produce a fix and the *ability* of the automatic repair technique. Also, they mention that there is no correlation between the test suites' *coverage* and the *ability* of the automatic repair technique, but the correlation exists between the test suites' *coverage* and the *quality* of the generated patches [86], [45].

Motwani et al. [86] conclude that the automatic program repair operations are inadequate in generating patches while the developer is required to add loops and/or new functions calls and /or to adjust the signatures. Therefore, the current article would like to suggest that there is a need for change in our *modus operandi*. The manner we approach the automatic program repair systems requires a shift in which it would further enable us to actually use them instead of a human developer in that position. We should remember that the motivation for moving toward full automation is to reduce the human impact and to transfer the responsibility of bug fixing from a human to a machine which is insofar not attained, but it is still an attainable goal.

## 5. Contextual analysis of the codes

As demonstrated above, most automatic program repair techniques fail when the complexity of the situation is high. The current article suggests that by mapping the differences and the location of the effects of those differences we may be able to better understand what specific path would lead us to 'full' automation. Diffraction method is often used to scrutinize several unseen angles and perspectives and to look for complexities and layers of the realities at hand [5]. The diffractive analysis is mostly used on qualitative data and theories where language has a pivotal role. Current articles suggest that the *codes;* specifically of bugs, should be considered as the context along with theories when applying the diffractive method on the subject of program repair. Furthermore, we suggest that the comparison of bugs should work as the comparison of contextual content. Therefore, two forms of comparison can be possible. Firstly, the comparison between two bugs of one program where different levels of automation is involved in order to understand the differences and where these differences have effects and secondly, the comparison between two similar bugs of two programs *if* different levels of automation is involved to understand the underlying reasons for the amount of human developer involvement. Current article suggests that in order to achieve the objective of this section, the *maintainability* of the codes should be compared (HELP! please correct me if my understanding is flawed).

Supplemental to what has been suggested insofar, it is interesting to see what other factors may impact the fix. In self-repair systems, it is found that if the fix is generated based on the human-made templates, they have a higher chance of getting the approval from the human supervisor [18]. The given fact magnifies the biases of human interaction in the fix generating process. Developers are more likely to approve the fixes that are based on templates they created in comparison to the fixes that are generated by atomic change [17]; which illustrate the gap between an entirely automatic process and the current process that includes human assistance. This is highly important since it underlines that there is a recognizable difference between generated fixes that is similar to what a human developer would produce and that of an automatic system. Mapping and approximating the differences while comparing the two may also give us important insights. Such insights may help us understand what steps should be taken to make the automatic program to generate a fix similar to what the human would have produced in the same situation.

## 6. Experimenting with the diffractive apparatus

Analyzing a phenomenon diffractively implies that the authors are not trying to spot cause and effect relations and do not view entities as traditional subject-object with individual agencies. According to Barad [5] the agency is not pre-existing separately and/or is not merely assigned to humans but it has emerged in the interaction between humans and artefacts. Indeed, the technology is embodied by certain artefacts such as computers and/or the effects of it on other materials. The primary intra-action relative to this article is the phenomenon that *becomes* between the human developer and the bug. This intra-action is *bug fixing*. It is meaningful because the relationship between a bug and a human determines it. However, in order to move toward full automation, we are required to manipulate this intra-action. For instance, in current theories whereby the program repair process is not fully automatic, but it is regarded in texts as if it is, the sociotechnical assemblage of the human developer and the bug fixing process (intra-action) is represented as one independent whole. It is how the current semi-automatic process is materialized. However, we need to disentangle this whole in order to improve the situation.

The essential element is to imitate the current intra-activity between the human developer and a bug (the debugging process) in a way that the same dynamic appears, and the agency occurs but without the human. In other words, as a part of the interaction between humans and the program we should notice that what is trying to be achieved through full automation is the elimination of the human by replacing it with capabilities in the program that characterize human cognitive capabilities in fixing bugs. A bug is an entity, an artefact that entangles with other bodies, e.g., the developer or; when the full automation is achieved, the automatic technology. Hence, the complete comprehension of the intra-action between a human developer and the bug is pivotal.

The human-handled aspects of debugging consist of analyzing the failure, spotting the cause, performing the fixing process and testing the debugged program to see if it works correctly without shortcomings [87], [88], [89]. Current manual activities are drastically expensive, which spotlights the lack of actual automation. These sets of intra-activities are insofar centralized on identifying entities with a higher chance of causing system failure [90], [91], [92]. When this identification is done automatically, the result of this automation is a report listing the possible entities that may cause failure which is sent to a tester to manually analyze if they are skeptical entities [90], [91], [92]. As it is illustrated almost all crucial aspects of this dynamic is done by a human developer. The medium of the automation here is mostly merely moderation. The agility can be solely attributed to the human developer since the automation level is minor.

This problem ensued a new set of *program repair techniques* [15], [16], [17], [48], [93], [94], [95], [31], [96], [32] in which the fix is also produced by automatic services and is later executed by the tester which reduces the amount of manual work. After years and years of research in this area, the software repair discipline thrived after the first search-based algorithms which could generate fixes automatically [97], [98]. On this level, where the program is facilitated by automatic techniques with agility, the intra-activity between the human developer and the program becomes more balanced. There exists a level of trust on this level regarding the accuracy of discernments done by the automatic programs. This elevated trust between computer and human opens the door for adopting new automatic cost-effective technologies.

Nonetheless, the debugging intra-action will impact other elements such as the financial aspects involved, employment, the industry specific narrative and etc. all intra-activities impacting the artefacts and humans involved from the beginning to end of a debugging process will be heavily influenced by reaching full automation. Specifically, if the human developer is eliminated from the process then the whole experience of bug fixing is altered. Furthermore, technology may change the way automation is materialized. The sociotechnical entanglements are invariably changing in the road to full automation. Current article suggests that the material practice of automatic debugging which involves the computer, various stages depended on the program at hand, the human developer, the environment where the interaction between the human developer and the bug occurs, narrative surrounding the automatic program repair in IT related fields of occupation and education and the financial consequences can be scrutinized as a *diffractive apparatus*.

This diffractive apparatus is of great importance since as it was imparted in the introduction, it takes up to 50% of the development process of a software production. Such immense financial consequence will be decreased if the human developer is eliminated from the diffractive apparatus. As it is postulated by Monperrus et al., [51], at least partial elimination of human developers is achieved if the program repair becomes *human-competitive*. In an exciting experiment, Monperrus et al. [99] proved that program repair is human-competitive. They created a bot named Repairnator which perpetually monitors bugs discovered in open-source software and tries to fix them automatically. They assigned two missions for Repairnator bot: a) to find the patch faster than a human developer and b) to be able to generate a fix that is correct-enough compared to a fix generated by a human in quality and readability [99]. Monperrus et al. [99] postulate that if the bot accomplishes the two mentioned missions, then program repair is human-competitive.

In order to conduct the experiment and avoid biases against machines, they created a profile with name and picture on GitHub for Repairnator. Within six months, Repairnator was able to generate five patches that got proved by the developer and were implemented. This experiment illustrates an excellent potential for future work in automatic repair systems [100], [99]. Although, we should move toward a point that the bots can generate the proper fixes as frequent as human developers. The second step is to overcome the obstacle of overfitting. Bots are unable to understand that a patch is incorrect because of overfitting which necessitates human interference [100], [99]. This experiment ensues two interpretations. First, it is highlighted that in circumstances where the program repair becomes human-competitive, while the elimination of human developers is promised, an additional intra-activity between the program repair and human supervisor will stay intact. However, if the human-competitive program repair is achieved regarding higher complexity faults, less or none, human supervision can be predicted in the long run. Although, the presence of a human supervisor is perhaps more cost-effective in all cases. Secondly, this experiment opens the door to the possibility of improving the human-competitiveness of the program repair by implementing machine learning algorithms.